\documentclass[prl,showpacs,twocolumn,letterpaper]{revtex4}
\usepackage{graphics}

\def\bea{\begin{eqnarray}}
\def\eea{\end{eqnarray}}
\def\ben{\begin{equation}}
\def\een{\end{equation}}
\def\benu{\begin{enumerate}}
\def\enu{\end{enumerate}}

\def\sss{\scriptscriptstyle\rm}

\def\1var{(\bx_1...\bx\N)}

\def\half{\frac{1}{2}}

\def\br{{\bf r}}

\def\bx{{\br t}}

\def\Hxc{_{\sss HXC}}

\def\N{_{\sss N}}

\def\sph_int{ {\int d^3 r}}

\begin{document}

\title{Measuring the kernel of time-dependent density functional
theory with X-ray absorption spectroscopy of $3d$ transition metals}
\author{A. Scherz}
\thanks{New address: SSRL, Stanford Linear Accelerator Center,
2575 Sand Hill Road, Menlo Park, California 94025, USA}
\author{E.K.U. Gross}
\author{H. Appel}
\author{C. Sorg}
\author{K. Baberschke}
\thanks{Corresponding author. Phone +49 30 838-52648, Fax +49 30 838-53646}
\email{bab@physik.fu-berlin.de}
\author{H. Wende}
\affiliation{Fachbereich Physik, Freie Universit\"{a}t Berlin, Arnimallee 14, D-14195 Berlin-Dahlem, Germany.}
\author{K. Burke}
\affiliation{Department of Chemistry and Chemical Biology, Rutgers University, 610 Taylor Rd, Piscataway, NJ 08854}
\date{October 28, 2005}
\begin{abstract} 
The $2p-3d$ core hole interaction in the $L_{2,3}$ absorption spectra of 
the $3d$ transition metals is treated within  
time-dependent density functional theory (TDDFT).
A simple three-level model explains the origin of the
strong deviations from the one-particle branching ratio and
yields matrix elements of the unknown
exchange-correlation (XC) kernel directly from experiment.
\end{abstract}


\maketitle


Ground-state density functional theory (DFT) 
is well-established
for atoms, molecules, and solids \cite{FNM03}. 
But ground-state DFT produces only a one-particle picture of the electronic
transitions in matter, neglecting interactions between excitations.
In the optical regime, time-dependent DFT (TDDFT) has enjoyed recent success
\cite{PGG96,reining,sole} in describing such dynamic
exchange-correlation (XC) effects.
The spectroscopic properties of matter in the X-ray regime are substantially governed
by dynamical many-body effects involving the creation of a localized core
hole \cite{zaanen85,schwitalla,shirley,corerehr,eriksson}.
While GW calculations and the
Bethe-Salpeter equation can be used \cite{shirley},
these are computationally demanding. The less
expensive methodology of TDDFT is now being 
developed for these effects \cite{corerehr}.

We analyze this approach to the X-ray absorption of
itinerant systems like the $L_{2,3}$ absorption of $3d$ transition metals (TMs),
i.e., exciting a photoelectron from the
localized $2p$ core states into the $3d$ band. 
$L_{2,3}$ X-ray absorption spectra (XAS), especially of early $3d$ TMs,
suffer from core-hole correlation effects \cite{zaanen85}.
Schwitalla and Ebert \cite{schwitalla} applied
TDDFT linear response theory 
to calculate the XAS of the
$3d$ TMs. Using a local approximation to the frequency-dependent XC kernel,
as proposed by Gross and Kohn \cite{gross_kohn},
they qualitatively reproduced the trend of the branching ratios.
However, whenever DFT is applied in a new regime, a difficult question
arises:  Are the existing functional approximations sufficiently accurate
in this new regime?   And how does one separate XC errors from those due to the
practical approximations needed for realistic calculations?
In Ref. \cite{corerehr}, the limitations of the
Gross-Kohn approximation for this problem are shown,
and a new approximation suggested.
But the true value of DFT is in constructing one XC approximation that
covers many situations, in order to build-in knowledge of the underlying
physics.  Other TDDFT approximations, such as Vignale-Kohn \cite{VK96},
would need to be
inserted into their codes in order to be tested.

Our approach here is different,
and is based on the  philosophy of Ref.~\cite{AGB03}.
That work examined the TDDFT
response when excitations are not strongly coupled to each other.
A useful series was developed in the strength
of the off-diagonal matrix elements, relative to the frequency shifts
induced by diagonal terms.  The leading term yields the single-pole
approximation (SPA)\cite{PGG96}, which has proven very useful in understanding
TDDFT corrections to the one-particle picture.  It even yields an immediate
estimate of the XC kernel, but only if excitations are well-separated,
a criterion rarely realized in practice \cite{AGB03,chelikowsky}.

\begin{figure}
  {\par\centering
  \resizebox*{0.85\columnwidth}{!}{\includegraphics{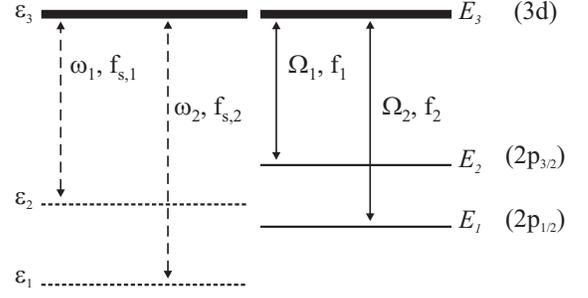}} \par}
  \caption{Schematic illustration of the DPA model. The model describes
  the shifts of the excitation energies (uncorrelated $\omega_i$ and 
  correlated $\Omega_i$) and the changes in corresponding oscillator
  strengths $f_i$ in the presence of an excited core hole.   
  \label{dpa}}
\end{figure}

However, the same philosophy applies to 
cases of two levels {\em strongly} coupled to one another,
but weakly coupled to the rest of the spectrum.
We call this the three-level or double-pole approximation (DPA), cf. Fig~\ref{dpa}.
Moreover, the $L_{2,3}$ absorption of $3d$ TMs provides
an ideal example of two transitions much closer to each other than the
rest of the spectrum.  With this in mind, we 
experimentally measured the branching
ratios and level splittings of the
$2p_{3/2}$($L_3$) and  $2p_{1/2}$($L_2$) core states, and now
{\em deduce} off-diagonal matrix elements
of the unknown XC kernel.
Because we can also compare with the one-particle Kohn-Sham (KS) spectrum,
we can also deduce the diagonal matrix elements.
The large deviation of branching ratios from their single-particle
values is due entirely to the effect of core-hole interaction
on spin-orbit coupling.
Thus the DPA to TDDFT explains the observed shifts and oscillator strengths,
and also provide benchmarks for future XC kernel approximations.
We believe this is the first experimental measurement of a matrix
element of the XC kernel of TDDFT.


Consider a system of electrons subject to a small
frequency-dependent perturbation.
By virtue of TDDFT the corresponding linear density-density response
function $\chi$ is related to the response function $\chi_s$ of 
non-interacting particles via the Dyson-type equation \cite{gross_kohn}
\begin{eqnarray}
\lefteqn{     \chi(\mathbf{r},\mathbf{r}^{\prime},\omega) = \chi_s(\mathbf{r},\mathbf{r}^{\prime},\omega)} \nonumber \\
               & &    + \int\mbox{d}^3x\int\mbox{d}^3x^\prime \chi_s(\mathbf{r},\mathbf{x},\omega)
                K(\mathbf{x},\mathbf{x}^{\prime},\omega)\chi(\mathbf{x}^\prime, \mathbf{r}^{\prime},\omega) .  
                \label{relatedResponse}
\end{eqnarray}
The kernel $K(\mathbf{r},\mathbf{r}^{\;\prime},\omega)$ consists of the bare Coulomb interaction and the 
frequency-dependent XC kernel  $f_{xc}(\mathbf{r},\mathbf{r}^{\;\prime},\omega)$:
\begin{equation}
                K(\mathbf{r},\mathbf{r}^{\prime},\omega)=
                \frac{e^2}{\left|\mathbf{r}-\mathbf{r}^{\prime}\right|}+
                f_{xc}(\mathbf{r},\mathbf{r}^{\prime},\omega) . 
                \label{Kernel}
\end{equation}
The exact XC kernel describes, among other many-body effects, the core-hole interaction
with the photoelectron. Neglecting $K$, the spectrum would reduce to the bare KS 
single particle spectrum represented by $\chi_s$.
In XAS, the deviations produced by $K$ are called core-hole
correlation effects. 
The response function $\chi_s$ is given in terms of the ground-state
KS orbitals $\varphi_j$ (spin-saturated) by
\begin{equation}
                \chi_s(\mathbf{r},\mathbf{r}^{\prime},\omega)=2\sum_{j,k}(n_j-n_k)
                \frac{\varphi^{\ast}_k(\mathbf{r})\varphi^{\ast}_j(\mathbf{r}^{\prime})
                \varphi_j(\mathbf{r})\varphi_k(\mathbf{r}^{\prime})}
                {\omega-\omega_{jk}+i\eta}
\label{chiS}
\end{equation}
where $n_j, n_k$ denote the Fermi-occupation factors and $\omega_{jk}$ are the KS
orbital-energy differences. For a single-particle transition $q$ ($q\equiv k\to j$)
define $\Phi_q(\mathbf{r}):=\varphi_k(\mathbf{r})\varphi_j^\ast(\mathbf{r})$.
The exact density-response function $\chi$ has poles at the true, correlated,
excitation energies $\Omega _j$, which can be found by
solving 
\cite{C96}:
\begin{equation}
\sum_{q'} {\tilde W}_{qq'} (\Omega_j)\ v_{q',j} = \Omega^2_j\ v_{q,j},
\label{mat}
\end{equation}
where 
the matrix is
\bea
{\tilde W}_{qq'}(\Omega)
&=& \omega^2_q \, \delta_{qq'} + 4 {\sqrt{\omega_q\, \omega_q'}}\,
K_{qq'} (\Omega),\nonumber\\
K_{qq'} (\Omega)&=&
\int d^3r\int d^3r'\, \Phi^*_q(\br)\,
K(\br\br'\Omega)\,\Phi_{q'}(\br').
\label{Mdef}
\eea
The eigenvectors yield the oscillator strengths\cite{C96} via
\ben
\label{oscgen}
f_j = \frac{2}{3}\, | \vec x^T\,S^{-\half}\, \vec v_j |^2,
\een
where
$S^{-\half}_{qq'} = \delta_{qq'}\; 
\sqrt{\omega_q}$ 
and $x_q$ is a column of the KS dipole matrix elements.
This eigenvalue problem rigorously determines the excitation spectrum of
the interacting system, but the quality of the results in
any practical calculation depends crucially on the 
approximation employed for the XC kernel.

In the $L_{2,3}$ XAS of the $3d$ TMs, 
the description of the electron core-hole interaction
may be simplified by the assumption that the relativistic spin-orbit 
coupling (SOC) in the $3d$ band states ($\sim 0.05$~eV)
is small compared to that of the core states (several eV)
and can be neglected.
This means that the oscillator strengths $f_j$ of these levels are
all about equal, as their KS orbitals are essentially identical.
Since, in this limit, the absorption area is proportional to the oscillator strength,
weighted statistically according to the manifold of the $j=3/2$ and
$j=1/2$ subshells, the branching ratio
of the KS system is $B_\mathrm{KS}=A_{3/2}/(A_{3/2}+A_{1/2})\equiv 2/3$,
where $A_j$ is the area under the peak of the $j$-th subshell.
(\emph{Ab initio} calculations of the $L_{2,3}$ XAS
without core-hole correlation effects based on the fully-relativistic
spin-polarized Korringa-Kohn-Rostoker band structure formalism
yield branching ratios very close to $B_\mathrm{KS}$ \cite{v-paper,xafs12}.)
Here we replace all dipole-allowed transitions $\omega_{jk}$ from a particular absorption edge
into the $3d$ band by a single particle transition, as illustrated in Fig. \ref{dpa}.

\begin{figure}
  {\par\centering
  \resizebox*{0.85\columnwidth}{!}{\includegraphics{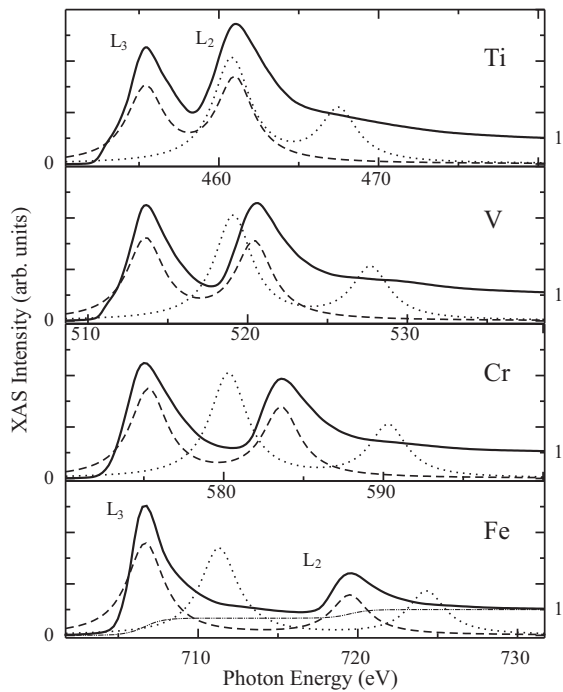}} \par}
  \caption{The experimental isotropic absorption spectra (solid line) at the $L_{2,3}$ edges
  are shown for the early $3d$ TMs Ti, V, and Cr versus Fe. The edge jumps are normalized
  to unity for direct comparison. The continuum in the experimental spectrum is
  simulated by a two-step function as shown for Fe (dashed-dotted line).
  The treatment of the core hole red-shifts the independent particle spectrum (dotted line)
  and changes the statistical branching ratio in the correlated spectrum (dashed line)
  as revealed by the DPA model.
  \label{spectra}}
\end{figure}

In Fig.~\ref{spectra}, we show our experimental isotropic XAS for the $3d$ TM
with almost empty $3d$ bands taken from Fe/TM/Fe sandwiches with
$\mathrm{TM}=\mathrm{Ti, V, Cr}$ and bulk-like Fe.
The data were recorded at the UE56-1/PGM 
beamline at BESSY
(for details, see Ref. \cite{xafs12}). 
The edge jumps are normalized to unity.
From these spectra and their absolute energy dependence,
the excitation energies $\Omega _{q=1}$ at
the $L_3$ edge and $\Omega_{q=2}$ and the $L_2$ edge are determined.
For the quantitative analysis of the branching ratio $B$, we very carefully determined the
$L_{2,3}$ absorption areas $A_j$. This determination has the advantage
that $B$ becomes independent of the
different $L_3$ and $L_2$ lifetime broadening and experimental resolution.
Note, that the proper experimental intensity is given by the area and not
by the height of the resonance. To determine the correct area of the $L_3$
and $L_2$ resonances the continuum contribution is removed
(e.g. gray line for Fe in Fig.~\ref{spectra} \cite{chenprl}). 
Since the $2p$ SOC decreases towards lower atomic numbers
the deconvolution is more complicated for
the early $3d$ TMs Ti, V, and Cr because of the strong $L_{2,3}$ overlap.
The areas have been fitted using the Fe absorption spectrum
as a background simulation underneath the $L_2$ edge. This fit appears 
justified, since the $L_2$ onsets for the early
$3d$ TMs follow systematically the energy dependence of the $L_3$ edge of Fe.
The experimental results are set out in Table~\ref{tab1}.
The energy resolution of the experimental spectra (Fig. \ref{spectra})
is approximately 0.5 eV. Consequently, the widths of a few eV in the 
$L_3$ and $L_2$ resonances are mostly due to lifetime effects.

In the case of Fe the $L_2$ absorption
is approximately half of the $L_3$ peak, in agreement with the KS prediction.
However, the branching ratios for the other $3d$ elements differ significantly
from this. In particular, Ti has an
$L_2$ peak that is even larger than its $L_3$ absorption. 
Thus, the experimental branching ratios cannot be interpreted in
terms of KS orbitals, suggesting strong electron core-hole interactions.
In the language of TDDFT, there must be significant off-diagonal matrix elements
in Casida's equations, describing
the influence of the electron core-hole interaction
on the $L_{2,3}$ XAS.
(If only diagonal elements are considered in Eq. (\ref{mat}),
the eigenvalues are shifted but
the eigenvectors are not rotated, and the oscillator strengths retain their
KS values \cite{AGB03}.)

However, a fully numerical solution of the equations is
not needed, as we know there are only two dominant transitions, so
the electron core-hole interaction
can be analyzed within the DPA model.
For only two transitions, we solve Eq. (\ref{mat}) exactly:
\bea
f_1&=&({\sqrt{f_{{\sss S},1}}}\; \cos(\theta/2) -
{\sqrt{f_{{\sss S},2}}}\; \sin(\theta/2))^2\nonumber\\
f_2&=&({\sqrt{f_{{\sss S},2}}}\; \cos(\theta/2) + 
{\sqrt{f_{{\sss S},1}}}\; \sin(\theta/2))^2
\label{osc}
\eea
for the interacting oscillator strengths $f_i$, where
\ben
\tan\theta=2\,W_{12}/(W_{22}-W_{11})
\label{tanth}
\een
is a mixing angle that represents the strength of the coupling between the
two transitions.  This yields:
\def\so{_{\sss S,1}}
\def\st{_{\sss S,2}}
\ben
\cos\theta =
\frac{4{\sqrt{f\so f\st q_1 q_2}}+(f\st-f\so)(q_2-q_1)}{(f\so+f\st)(q_1+q_2)}
\label{thetafull}
\een
where $f_{\sss S,i}$ denotes a Kohn-Sham oscillator strength and
$q_1=B p_2$, $q_2 = (1-B) p_1$, where 
$p_i$ the multiplicity of the initial state of transition $i$.
In our case, setting $f_{{\sss S},1}=f_{{\sss S},2}=f_0$, 
Eq. (\ref{osc}) simplifies to
\ben
f_1 = f_0\; (1 - \sin\theta ),~~~~~f_2 = f_0\; (1 + \sin\theta ),
\een
and since the absorption depends only on the oscillator strengths, weighted
statistically,  Eq. (\ref{thetafull}) yields 
\ben
\sin\theta = (2 - 3 B)/(2 - B).
\een
The branching ratio directly determines the mixing angle!
We can even extract directly the off-diagonal matrix element of the
Hartree-XC kernel itself. Inserting the matrix elements
into Eq. (\ref{tanth}), and neglecting all small differences (e.g., between KS and exact
transition frequencies) which are a few eV in several hundred,
\ben
K_{12} = \sin\theta\; \Delta \Omega /4 \; .
\label{Koff}
\een
Thus knowledge of both the branching ratio and the level splitting
$\Delta\Omega=\Omega_2 - \Omega_1 $
yields an experimental determination of the off-diagonal matrix element
of the XC kernel. 
Lastly, given the ground-state KS energy levels,
we can also recover the diagonal elements:
\ben
K_{jj}=(\Omega_{1}+\Omega_{2})/4 + (-1)^j \Delta\Omega\; \cos\theta\; /4 - \omega_j/2\; \quad j=1,2.
\label{k11k22}
\een
These results are listed in Table \ref{tab1}, the corresponding theoretical DPA
spectra are presented in Fig.~\ref{spectra}.

This analysis yields a simple interpretation of the observed
spectra.  First, imagine there was no SOC.  Then there
would be a single $p$-level, and SPA applies.  The diagonal matrix
element of $f\Hxc$ is the well-known core-hole interaction that shifts
the transition frequency from its KS value. 
In the presence of spin-orbit splitting, both levels are shifted by
similar amounts, about 5-7 eV.

Much more importantly, a  new effect occurs, which is that the
core-hole interaction {\em couples} the two KS transitions together,
altering their branching ratio.
The much smaller {\em off-diagonal}
core-hole interaction (about 1/2 eV) produces the large deviation from
the single-particle branching ratio.  Although the matrix elements are about
5 times smaller than their diagonal counterparts, in Ti they reverse the
relative sizes of the peaks.  This effect can be thought of
as simple level (or in this case, transition) repulsion, 
as the two transitions near
one another.  Eq. (\ref{Koff}) shows that the true measure of coupling is
$4 |K_{12}|/\Delta \Omega$, which is growing from Fe to Ti only because
the $2p$ SOC $\Delta\Omega$
is shrinking, not because of increased interaction.
We also note that SPA \cite{PGG96} (which can be recovered in
all results by setting $\theta=0$) yields $2K_{jj}=\Omega_j-\omega_j$ and
is highly accurate for the {\em diagonal} elements.   Thus the shifts
are simply interpreted as diagonals of $K$, while the
branching ratios are a sensitive determinant of off-diagonal elements.
Lastly, for
very small splitting ($\Delta \Omega \ll 4 K_{12}$),
$\sin\theta \to 1$ and, from Eq. (\ref{osc}), all weight goes into
the $L_2$ peak.  
From Eqs.(\ref{Koff}-\ref{k11k22}):
\ben
\Delta\Omega = 2 \sqrt{ (\Delta\omega/2 + K_{22}-K_{11})^2 + 4K_{12}^2}\; .
\een
Thus $4|K_{12}|$ is the
minimum level splitting, and occurs with the $L_2$ peak much larger than $L_3$.

The success of DPA shows that very little effort beyond a ground-state
DFT calculation is needed to compute these spectra in TDDFT.
One only needs to integrate a given
approximation to the XC kernel for the two diagonal matrix elements, and one off-diagonal.
Furthermore, applying our analysis in reverse to the calculated ALDA and RPA results of
Fig. 1 of Ref. \cite{corerehr}, we find that the success of the suggested approximation
(which implies using ALDA for diagonal elements, and neglecting
XC for the off-diagonal elements, i.e., RPA) also implies that ALDA is accurate for the peak
shifts alone, while RPA is accurate for the branching ratio predicted by Eq. (\ref{Koff}),
{\em once} the experimental shift is used.

Why is DPA justified for these systems?  There are two clear
sources of error.  On the one hand, there are transitions to other levels
to consider, but these are well-known to have small effects \cite{W04}.
For example, the transitions to $4s$ have much smaller oscillator
strengths, while other transitions are included in the background, which has
been subtracted.  Of greater concern might be the lack of resolution of
the (slightly) split $3d$ levels, which yield a 9$\times$9 Casida matrix problem
of allowed transitions.  However, just as DPA reduces to the single-pole
approximation of Ref. \cite{AGB03} when levels are
too close to be resolved \cite{AGB06}, we expect the full solution of the
9$\times$9 problem to collapse to the DPA results
when the individual $d$-levels cannot be resolved.  This will only be true for
the early TM's, in which most of the  $3d$-levels are unoccupied.


\begin{table}
\caption{\label{tab1} 
Excitation energies in eV obtained from KS calculations
($\omega_i^{KS}$) and from experiment
($\Omega_i$), experimental branching ratio $B $
and matrix elements $K_{ij}$.
The experimental error of $\Omega_i$ is below $10^{-3}$,
the one of $B$ in the order of 1~\%.}
\begin{ruledtabular}
\begin{center}
  \begin{tabular}{lrrrrrrrr} 
	$3d$ TM & $\omega_1^{KS}$ & $\omega_2^{KS}$ & $\Omega_1$ & $\Omega_2$ & $B$ & $K_{11}$ &
	$K_{22}$ & $K_{12}$ \\ \hline
22 Ti &    460.8 &    467.5 &    455.4 &    461.0 &     0.47 &    -2.57 &    -3.34 &     0.54 \\ 
23 V  &    519.1 &    527.7 &    513.6 &    520.4 &     0.51 &    -2.65 &    -3.73 &     0.54 \\ 
24 Cr &    580.3 &    590.3 &    575.1 &    583.6 &     0.56 &    -2.55 &    -3.40 &     0.47 \\ 
26 Fe &    711.3 &    724.6 &    706.7 &    719.5 &     0.70 &    -2.29 &    -2.55 &    -0.25 \\ 
  \end{tabular}
  \end{center}
\end{ruledtabular}
\end{table} 

In summary, we have used TDDFT to understand the XAS of
$3d$ transition metals by deriving a double-pole approximation. The main
features observed in the experiments can easily be explained by assuming that the spectrum
is dominated by two strongly coupled poles via the $2p-3d$ core hole interaction.
This shows that, for the beginning
of the $3d$ series, the reduced $2p$-SOC is responsible for the strong variation
of the branching ratio, not strong interactions between the transitions.
Our analysis does not replace a full TDDFT calculation of X-ray
absorption spectra. Rather, for the very specific case of spectral
regions dominated by two poles it provides, on the one hand,
a transparent picture of the changes of
spectral weights in particular for the early $3d$ TMs, and on the other, a
straight-forward route to testing approximate XC kernels against experimental data.

Discussions with J.~J. Rehr and U. Diebold are acknowledged.
We also thank the authors of Ref.~\cite{corerehr} for their KS eigenvalues.
The work was supported by BMBF (05 KS4 KEB/5) and DFG, Sfb 290. Partial
financial support by the EXC!TING Research and Training Network of the EU and the NANOQUANTA network
of excellence is gratefully acknowledged.
KB thanks the US DOE
(DE-FG02-01ER45928) and NSF (CHE-0355405).

\end{document}